\documentstyle[amssymb,thmsa,12pt,sw20lart]{article}
\input{tcilatex}
\begin{document}

\topmargin 0pt \oddsidemargin 5mm

\setcounter{page}{1}

\hspace{8cm}{} \vspace{2cm}

\begin{center}
{\large {DICLUSTER STOPPING IN A DEGENERATE ELECTRON GAS}}\\[0pt]

Hrachya B. Nersisyan$^{(1)}$ and Amal K. Das$^{(2)}$\\[0pt]
\vspace{1cm} $^{(1)}$ {\em Division of Theoretical Physics, Institute of
Radiophysics and Electronics, Alikhanian Brothers St. 2, Ashtarak-2, 378410,
Republic of Armenia}\footnote{%
E-mail: hrachya@irphe.am}
\end{center}

$^{(2)}$ {\em Department of Physics, Dalhousie University, Halifax, Nova
Scotia B3H 3J5, Canada}

\vspace {5mm} \centerline{{\bf{Abstract}}}

In this paper we report on our theoretical studies of various aspects of the
correlated stopping power of two point-like ions (a dicluster) moving in
close but variable vicinity of each other in some metallic target materials
the latter being modelled by a degenerate electron gas with appropriate
densities. Within the linear response theory we have made a comprehensive
investigation of correlated stopping power, vicinage function and related
quantities for a diproton cluster in two metallic targets, aluminum and
copper, and present detailed and comparative results for three
approximations to the electron gas dielectric function, namely the
plasmon-pole approximation without and with dispersion as well as with the
random phase approximation. The results are also compared, wherever
applicable, with those for an individual projectile.

\newpage

\section{INTRODUCTION}

The stopping of energetic charged particles in a target material is a
problem of long-standing theoretical and experimental interest. Early
pioneering theoretical work by Bohr who had a lifelong interest in this
problem was followed by Bethe and others. There is by now an extensive
literature on this topic. We refer the reader to two recent review articles
[1, 2]. The problem in its generality is rather complex. However simplified
theoretical models have been applied with considerable success in explaining
experimental data. Assuming a weak coupling between the energetic particle
and a target material, specially a metal which is usually approximated by a
degenerate electron gas, a detailed theoretical model has emerged through
the works of Bohm, Pines, Lindhard, Ritchie and other authors [3-5].

A comprehensive treatment of the quantities related to inelastic
particle-solid and particle-plasma interactions, e.g. scattering rates and
differential and total mean free paths and energy losses, can be formulated
in terms of the dielectric response function obtained from the electron gas
model. The results have important applications in astrophysics [6, 7] and
radiation and solid-state physics [8-11], and more recently, in studies of
energy deposition by ion beams in plasma fusion targets [12-16].

One can think of several situations in which the projectile beam ions may be
closely spaced so that their stopping is influenced by their mutual
interactions [14, 15] and thus differs from the stopping of charged
particles whose dynamics is independent of each other. This can happen, for
example, in the case of very high density beams, or more realistically, when
cluster ions are to be used instead of standard ion beams [16].

The stopping of uncorrelated or independent charged particles in a
degenerate electron gas (DEG) has been extensively studied in the literature
(see, for example, [3-9, 13] and other references therein). These studies
have been done mostly within the linear response formalism and for the
projectile velocity $V$ comparable or greater than $v_F$, the electron Fermi
velocity.

The objective of this paper is to make a study of the stopping power (SP) of
correlated charged particles in a DEG. The simplest and yet physically
relevant case is the SP of an ion pair (a dicluster). We report on a
comprehensive investigation, which is mostly numerical, of various aspects
of a dicluster stopping in a DEG. Previous theoretical works have considered
this problem within the linear response theory [9-11] and in a simple
plasmon-pole approximation. In our study, again in the linear response
formalism, we have used both the plasmon-pole approximation (PPA) as well as
the full Lindhard expression for the random phase approximation (RPA) in the
DEG dielectric function. As in earlier studies we consider a diproton
cluster as a projectile and compare our theoretical results with those of
Basbas and Ritchie [10] who used PPA and with those obtained in RPA.
Whenever applicable, the results are also compared with those for an
independent projectile (e.g. those due to Yakolev and Kotel'nikov [7]). No
RPA results for a diproton cluster SP and related aspects have previously
been reported in the literature.

\section{STOPPING POWER}

Let us consider an external charge with distribution $\rho _{{\rm ext}}({\bf %
r},t)=Q_{{\rm ext}}({\bf r}-{\bf V}t)$ moving with velocity ${\bf V}$ in a
medium characterized by the longitudinal dielectric function $\varepsilon
(k,\omega )$. Within the linear response theory and in the Born
approximation the scalar electric potential $\varphi ({\bf r},t)$ due to
this external charge screened by the medium is given by [1]

\begin{equation}
\varphi ({\bf r},t)=4\pi \int d{\bf k}Q_{{\rm ext}}({\bf k})\frac{\exp
\left[ i{\bf k}({\bf r}-{\bf V}t)\right] }{k^2\varepsilon (k,{\bf kV})}{\bf ,%
}
\end{equation}
where $Q_{{\rm ext}}({\bf k})$ is the Fourier transform of function $Q_{{\rm %
ext}}({\bf r})$.

The stopping power which is the energy loss of the external charge regarded
as a projectile, per unit path length in the medium regarded as a target
material, can be calculated from the force acting on the charge. The latter
is related to the induced electric field ${\bf E}_{{\rm ind}}$ in the
medium. For a three-dimensional medium we have, for the SP,

\begin{equation}
S\equiv -\int d{\bf r}Q_{{\rm ext}}({\bf r}-{\bf V}t)\frac{{\bf V}}V{\bf E}_{%
{\rm ind}}({\bf r},t)=\frac{2(2\pi )^4}V\int d{\bf k}\left| Q_{{\rm ext}}(%
{\bf k})\right| ^2\frac{{\bf kV}}{k^2} {\rm Im} \frac{-1}{\varepsilon (k,%
{\bf kV})}{\bf .}
\end{equation}

Eq. (2) is applicable to any external charge distribution. We shall apply it
to a cluster of two point-like ions having charges $Z_1e$ and $Z_2e$
separated by a variable distance ${\bf R}$. For this dicluster

\begin{equation}
\left| Q_{{\rm ext}}({\bf k})\right| ^2=\frac{e^2}{(2\pi )^6}\left[
Z_1^2+Z_2^2+2Z_1Z_2\cos \left( {\bf kR}\right) \right] .
\end{equation}

Then the stopping power of a dicluster may be written as

\begin{equation}
S=\left( Z_1^2+Z_2^2\right) S_{{\rm ind}}(V)+2Z_1Z_2S_{{\rm corr}}({\bf R}%
,V),
\end{equation}
where $S_{{\rm ind}}(V)$ and $S_{{\rm corr}}({\bf R},V)$ stand for
individual and correlated SP, respectively. From Eqs. (2) and (3)

\begin{equation}
S_{{\rm ind}}(V)=\frac{2e^2}{\pi V^2}\int_0^\infty \frac{dk}k\int_0^{kV}{\rm %
Im} \frac{-1}{\varepsilon (k,\omega )}\omega d\omega ,
\end{equation}

\begin{eqnarray}
S_{{\rm corr}}({\bf R},V) &=&\frac{2e^2}{\pi V^2}\int_0^\infty \frac{dk}%
k\int_0^{kV}{\rm Im}\frac{-1}{\varepsilon (k,\omega )}\omega d\omega  \\
&&\times \cos \left( \frac \omega VR\cos \vartheta \right) J_0\left( R\sin
\vartheta \sqrt{k^2-\frac{\omega ^2}{V^2}}\right) .  \nonumber
\end{eqnarray}
$J_0(x)$ is the Bessel function of the first kind and zero order and $%
\vartheta $ is the angle between the interionic separation vector ${\bf R}$
and the velocity vector ${\bf V}$. Eqs. (5) and (6) can also be obtained
from the linearized Vlasov-Poisson equations for a two-ion projectile
system, as has been done by Avanzo et al. [14]. In their study the target
material is a dense classical electron gas.

We note that there are two contributions to the SP of a two-ion cluster. The
first one, given by the first term in Eq. (4), is the uncorrelated particle
contribution and represents the energy loss of the individual projectiles
due to their coupling with the target electron gas. The second contribution,
the second term in Eq. (4), arises due to a correlated motion of the two
ions through a resonant interaction with the excitations of the electron
gas. Both terms are responsible for an irreversible energy transfer from the
two-ion projectile system to the target electron gas.

In many experimental situations, clusters are formed with random
orientations of ${\bf R}$. A correlated stopping power appropriate to this
situation may be obtained by carrying out a spherical average over ${\bf R}$
of the $S_{{\rm corr}}({\bf R},V)$ in Eq. (4). We find

\begin{equation}
\overline {S}_{{\rm corr}}(R,V)=\frac{2e^2}{\pi R V^2}\int_0^\infty \frac{dk%
}{k^2} \sin (kR) \int_0^{kV} {\rm Im} \frac{-1}{\varepsilon (k,\omega )}%
\omega d\omega.
\end{equation}

One may consider an interference or vicinage function, which is a measure of
the separation of single-particle contribution from its correlated
counter-part, to the stopping power. This function is defined as [10, 14, 15]

\begin{equation}
\Gamma ({\bf R},V)=\frac{S_{{\rm corr}}({\bf R},V)}{S_{{\rm ind}}(V)}.
\end{equation}
Eq. (4) can then be put in the form

\begin{equation}
S=\left( Z_1^2+Z_2^2\right) S_{{\rm ind}}(V)\left[ 1+\frac{2Z_1Z_2}{%
Z_1^2+Z_2^2}\Gamma ({\bf R},V)\right] .
\end{equation}
$\Gamma ({\bf R},V)$ describes the strength of correlation effects with
respect to an uncorrelated situation. The vicinage function becomes equal to
unity as $R\rightarrow 0$ when the two ions coalesce into a single entity,
and goes to zero as $R\rightarrow \infty $ when the two ions are totally
uncorrelated.

\section{THEORETICAL CALCULATIONS OF SP}

The key ingredient in the calculation of stopping power, as outlined in
Section II, is the linear response function $\varepsilon (k,\omega )$ of the
target material. The latter is modelled in most cases by a dense electron
gas neutralized by a positive background - the so-called jellium model. The
effect of lattice is not included except perhaps through an effective mass
of the electrons. This is a reasonable model for target materials like Cu
and Al. $\varepsilon (k,\omega )$ for a dense, and hence a degenerate
electron gas, has been calculated in various approximations in the
literature. Two of them are: (A) Plasmon-Pole approximation (PPA), and (B)
the full random phase approximation (RPA). Actually the plasmon-pole
approximation is a simplification of the RPA response function. We shall
consider SP in these two approximations.

\subsection{SP in PPA without plasmon dispersion}

Here we consider the simplest model of the dielectric function of a jellium.
In Ref. [10] (see also the Ref. [15]) a plasmon-pole approximation to $%
\varepsilon (k,\omega )$ for an electron gas was used for calculation of
dicluster SP. In order to get easily obtainable analytical results, Basbas
and Ritchie [10] employ a simplified form which exhibits collective and
single-particle effects

\begin{equation}
{\rm Im}\frac{-1}{\varepsilon \left( k,\omega \right) }=\frac{\pi \omega _p^2%
}{2\omega }\left[ H\left( k_c-k\right) \delta \left( \omega -\omega
_p\right) +H\left( k-k_c\right) \delta \left( \omega -\frac{\hbar k^2}{2m}%
\right) \right] ,
\end{equation}
where $H(x)$ is the Heaviside unit-step function, $\hbar \omega _p$ is the
plasma energy of the electron gas and the choice $k_c=\left( 2m\omega
_p/\hbar \right) ^{1/2}$ allows the two $\delta $ functions in Eq. (10) to
coincide at $k=k_c$ in the $k$-$\omega $ plane.

The first term in Eq. (10) describes the response due to nondispersive
plasmon excitation in the region $k<k_c$, while the second term describes
free-electron recoil in the range $k>k_c$ (single-particle excitations).
This approximate function satisfies the sum rule

\begin{equation}
\int_0^\infty {\rm Im}\frac{-1}{\varepsilon (k,\omega )}\omega d\omega =%
\frac{\pi \omega _p^2}2
\end{equation}
for all values of $k$.

In this approximation if $V>\left( \hbar \omega _p/2m\right) ^{1/2}\equiv
V_p $

\begin{equation}
S_{{\rm ind}}(\lambda )=\frac{e^2\omega _p^2}{V^2}\ln \left( \frac{2mV^2}{%
\hbar \omega _p}\right) =\frac{\Sigma _0}{\pi \chi ^2\lambda ^2}\ln \left( 
\frac{\lambda ^2\sqrt{3}}\chi \right) ,
\end{equation}
where $\lambda =V/v_F$, $\chi ^2=1/\pi k_Fa_0=\left( 4/9\pi ^4\right)
^{1/3}r_s$; $r_s=\left( 3/4\pi n_0a_0^3\right) ^{1/3}$, $n_0$ is the
electron gas density and $a_0=0.53\times 10^{-8}$cm is the Bohr radius. $k_F$
is the Fermi wave number of the target electrons and $\Sigma _0=2.18$
GeV/cm. In our calculations $\chi $ (or $r_s$) serves as a measure of
electron density. The result in Eq.(12) agrees exactly with the Bethe SP
formula, except that the plasmon energy of the electron gas $\hbar \omega _p$
appears instead of the usual mean atomic excitation energy. Eq. (12)
represents the contribution of valence/conduction electrons in a solid to
the stopping of an ion.

Using Eq. (10) in Eq. (6), in the high-velocity limit $V>V_p$ (or $\lambda
^2>\chi /\sqrt{3}\equiv \lambda _0^2$) one finds

\begin{eqnarray}
S_{{\rm corr}}(R,\vartheta ,\lambda ) &=&\frac{\Sigma _0}{\pi \chi ^2\lambda
^2}\left\{ \cos \left( \frac{2\chi }{\lambda \sqrt{3}}k_FR\cos \vartheta
\right) \right.  \\
&&\times \int_1^{\lambda /\lambda _0}\frac{dx}xJ_0\left( \frac{2\chi }{%
\lambda \sqrt{3}}k_FR\sin \vartheta \sqrt{x^2-1}\right) +  \nonumber \\
&&\left. \int_{2\lambda _0}^{2\lambda }\frac{dx}x\cos \left( \frac{x^2}{%
2\lambda }k_FR\cos \vartheta \right) J_0\left( xk_FR\sin \vartheta \sqrt{1-%
\frac{x^2}{4\lambda ^2}}\right) \right\} .  \nonumber
\end{eqnarray}

If one ion trails directly behind the other ($\vartheta =0$) from Eq. (13)
we find

\begin{eqnarray}
S_{{\rm corr}}(R,0,\lambda ) &=&\frac{\Sigma _0}{2\pi \chi ^2\lambda ^2}%
\left\{ \cos \left( \frac{2\chi }{\lambda \sqrt{3}}k_FR\right) \ln \left( 
\frac{\lambda ^2\sqrt{3}}\chi \right) +\right.  \\
&&\left. {\rm ci}\left( 2\lambda k_FR\right) -{\rm ci}\left( \frac{2\chi }{%
\lambda \sqrt{3}}k_FR\right) \right\} ,  \nonumber
\end{eqnarray}
where ${\rm ci}\left( z\right) $ is the integral cosin function

\begin{equation}
{\rm ci}\left(z\right) = -\int_z^{\infty} dx\frac{\cos x}{x}.
\end{equation}

One sees a characteristic oscillatory behavior for large interionic distance 
$R$. As discussed in [16], fluctuations in the stopping power of a medium
for a cluster as separation increases are due to electron density variation
in the wake of the leading ion. The wavelength of these fluctuations is $%
\sim 2\pi V/\omega _p$ for high-velocity projectiles.

In the case of randomly oriented clusters from Eq. (7) we find

\begin{equation}
\overline{S}_{{\rm corr}}(R,\lambda )=\frac{\Sigma _0}{\pi \chi ^2\lambda ^2}%
\left[ {\rm si_2}\left( \frac{2\chi }{\lambda \sqrt{3}}\left( k_FR\right)
\right) -{\rm si_2}\left( 2\lambda \left( k_FR\right) \right) \right] ,
\end{equation}
where

\begin{equation}
{\rm si_2}\left(z\right) = \int_z^{\infty} dx\frac{\sin x}{x^2} = \frac{%
\sin(z)}{z} - {\rm ci}(z).
\end{equation}

\subsection{SP in PPA with plasmon dispersion}

Plasmons without dispersion are an idealization. In real systems plasmons
are expected to undergo a dispersion leading to a $\omega (k)$. The actual
dispersion ( in RPA) can be obtained from the linear response function (see
Sec. 3.3). Here we shall utilize a dispersion which is valid for small and
intermediate values of the wave vector ${\bf k}$. Consequently we write

\begin{equation}
{\rm Im}\frac{-1}{\varepsilon \left( k,\omega \right) }=\frac{\pi \omega _p^2%
}{2\omega }\delta \left( \omega -\Omega (k)\right) ,
\end{equation}
where the dispersion is given by

\begin{equation}
\Omega ^2(k)=\omega _p^2+\frac 35k^2v_F^2+\frac{\hbar ^2k^4}{4m^2}.
\end{equation}
In this approximation when

\begin{equation}
V>\left( \frac 35v_F^2+\frac{\hbar \omega _p}m\right) ^{1/2}\equiv V_0
\end{equation}
we have, for ISP and CSP,

\begin{equation}
S_{{\rm ind}}(\lambda )=\frac{\Sigma _0}{\pi \chi ^2\lambda ^2}\ln \frac{%
\lambda ^2-3/5+\sqrt{\left( \lambda ^2-3/5\right) ^2-4\chi ^2/3}}{2\chi /%
\sqrt{3}},
\end{equation}

\begin{equation}
S_{{\rm corr}}(R,\vartheta ,\lambda )=\frac{\Sigma _0}{\pi \chi ^2\lambda ^2}%
\int_{x_{-}(\lambda )}^{x_{+}(\lambda )}\frac{dx}x\cos \left( \frac{\phi
_1(x)}{2\lambda }k_FR\cos \vartheta \right) J_0\left( \frac{\phi _2(x)}{%
2\lambda }k_FR\sin \vartheta \right) .
\end{equation}
Here $k_F$ is the Fermi wave number, and

\begin{equation}
\phi _1(x)=\sqrt{x^4+\frac{12}5x^2+16\chi ^2/3},
\end{equation}

\begin{equation}
\phi _2(x)=\sqrt{4\left( \lambda ^2-\frac 35\right) x^2-\left( x^4+16\chi
^2/3\right) },
\end{equation}

\begin{equation}
x_{\pm }(\lambda )=\sqrt{2\left[ \lambda ^2-\frac 35\pm \sqrt{\left( \lambda
^2-\frac 35\right) ^2-\frac{4\chi ^2}3}\right] }.
\end{equation}
In the case of randomly oriented clusters we find

\begin{equation}
\overline{S}_{{\rm corr}}(R,\lambda )=\frac{\Sigma _0}{\pi \chi ^2\lambda ^2}%
\left[ {\rm si_2}\left( k_FRx_{-}(\lambda )\right) -{\rm si_2}\left(
k_FRx_{+}(\lambda )\right) \right] .
\end{equation}

\subsection{Stopping power in RPA}

Now we will derive the analytical expressions for the SP of a dicluster in a
fully degenerate ($T=0$) electron gas. For this purpose we use the exact RPA
dielectric response function obtained by Lindhard [4]

\begin{equation}
\varepsilon (z,u)=1+\frac{\chi ^2}{z^2}\left[ f_1(z,u)+if_2(z,u)\right] ,
\end{equation}
where

\begin{equation}
f_1(z,u)=\frac 12-\frac 1{8z}\left( U_{+}^2-1\right) \ln \left| \frac{U_{+}+1%
}{U_{+}-1}\right| +\frac 1{8z}\left( U_{-}^2-1\right) \ln \left| \frac{%
U_{-}+1}{U_{-}-1}\right| ,
\end{equation}

\begin{equation}
f_2(z,u)=\left\{ 
\begin{array}{l}
\frac \pi {8z}\left( 1-\left( u-z\right) ^2\right) ;\quad |u-1|\leqslant
z\leqslant u+1 \\ 
0;\quad 0\leqslant z\leqslant u-1, \\ 
0;\quad z\geqslant u+1, \\ 
\frac 12\pi u;\quad 0\leqslant z\leqslant 1-u
\end{array}
\right. .
\end{equation}
Here, as in Refs. [6, 7, 13, 15], we have introduced the following notations 
$z=k/2k_F$, $u=\omega /kv_F$, $U_{\pm }=u\pm z$. With these notations Eqs.
(5), (6) and (7) read

\begin{equation}
S_{{\rm ind}}(\lambda )=\frac{6\Sigma _0}{\pi ^2\chi ^2\lambda ^2}%
\int_0^\infty z^3dz\int_0^\lambda \frac{f_2(z,u)udu}{\left[ z^2+\chi
^2f_1(z,u)\right] ^2+\chi ^4f_2^2(z,u)},
\end{equation}

\begin{eqnarray}
S_{{\rm corr}}(\lambda ,R,\vartheta ) &=&\frac{6\Sigma _0}{\pi ^2\chi
^2\lambda ^2}\int_0^\infty z^3dz\int_0^\lambda \frac{f_2(z,u)udu}{\left[
z^2+\chi ^2f_1(z,u)\right] ^2+\chi ^4f_2^2(z,u)}  \nonumber \\
&&\times \cos \left( 2\frac{zu}\lambda k_FR\cos \vartheta \right) J_0\left(
2zk_FR\sin \vartheta \sqrt{1-\frac{u^2}{\lambda ^2}}\right) ,
\end{eqnarray}

\begin{equation}
\overline{S}_{{\rm corr}}(R,\lambda )=\frac{3\Sigma _0}{\pi \lambda ^2}\frac{%
a_0}R\int_0^\infty \sin \left( 2k_FRz\right) z^2dz\int_0^\lambda \frac{%
f_2(z,u)udu}{\left[ z^2+\chi ^2f_1(z,u)\right] ^2+\chi ^4f_2^2(z,u)}.
\end{equation}

In order to evaluate the integrals by $z$ in Eqs. (30)-(32) at $\lambda <1$
(in the low-velocity limit) we split the integration region into two domain: 
$0\leqslant z\leqslant 1-u$ and $1-u\leqslant z\leqslant 1+u$, where ${\rm {%
Im}}$ $\varepsilon \sim f_2\neq 0$. However at $\lambda >1$ (in
high-velocity limit) we need to take into account the region $1\leqslant
u\leqslant \lambda $, $0\leqslant z\leqslant u-1$, where $f_2$ may vanish.
The integration in this region includes the excitation of collective plasma
modes (plasmons) by fast charged particles. Consequently, although $f_2=0$
the integrals in this region are not equal to zero. A calculation of the
collective part of SP is facilitated if we use the following known expression

\begin{eqnarray}
\frac{\chi ^2f_2(z,u)}{\left[ z^2+\chi ^2f_1(z,u)\right] ^2+\chi ^4f_2^2(z,u)%
} &\rightarrow &\pi \delta \left( z^2+\chi ^2f_1(z,u)\right) = \\
&=&\pi \frac{\delta \left( z-z_r(\chi ,u)\right) }{\left| 2z+\chi ^2\frac{%
\partial f_1(z,u)}{\partial z}\right| _{z=z_r(\chi ,u)}},  \nonumber
\end{eqnarray}
where $z_r(\chi ,u)$ is the solution of the dispersion equation $\varepsilon
(k,\omega )=0$ in variables $z$ and $u$.

Figure 19 shows the solution $z_r(\chi ,u)$ for various values of $\chi $
(solid line, $\chi =0.5$; dashed line, $\chi =0.15$; dotted line, $\chi
=0.05 $). It may be noted that the integration domain $0\leqslant u\leqslant
\lambda $, $z>u+1$, where $f_2=0$, does not contain the dispersion curve $%
z_r(\chi ,u)$ calculated for metallic densities $\chi \sim 0.5$ ($r_s\sim 2$%
). Consequently the SP in this region of variables $z$ and $u$ vanishes and
there is no plasmon excitation.

Let us consider the low-velocity limit ($V\ll v_F$) of Eqs. (30) and (31).
In this limit one can obtain simpler expressions for SP. From Eqs. (30) and
(31) we have

\begin{equation}
S_{{\rm ind}}(\lambda )\simeq \frac{\Sigma _0}{\pi \chi ^2}\lambda \int_0^1%
\frac{z^3dz}{\left[ z^2+\chi ^2f(z)\right] ^2},
\end{equation}

\begin{equation}
S_{{\rm corr}}(\lambda ,R,\vartheta )\simeq \frac{3\Sigma _0}{2\pi \chi ^2}%
\lambda \int_0^1\frac{z^3dz}{\left[ z^2+\chi ^2f(z)\right] ^2}\left[ \Phi
_1(zk_FR)+\Phi _2(zk_FR)\sin ^2\vartheta \right] ,
\end{equation}
where

\begin{equation}
\Phi _1(\xi )=\frac 1{\xi ^3}\left[ \left( \xi ^2-\frac 12\right) \sin
\left( 2\xi \right) +\xi \cos \left( 2\xi \right) \right] ,
\end{equation}

\begin{equation}
\Phi _2(\xi )=\frac 1{\xi ^3}\left[ \left( \frac 34-\xi ^2\right) \sin
\left( 2\xi \right) -\frac 32\xi \cos \left( 2\xi \right) \right] ,
\end{equation}

\begin{equation}
f(z)\equiv f_1(z,0)=\frac 12+\frac{1-z^2}{4z}\ln \frac{1+z}{1-z}.
\end{equation}

From Eqs. (36) and (37) it follows that $\Phi _1(\xi )\rightarrow 2/3$, $%
\Phi _2(\xi )\rightarrow 0$ at $\xi \rightarrow 0$ and consequently, as
expected, $S_{{\rm corr}}(\lambda ,R,\vartheta )\rightarrow S_{{\rm ind}%
}(\lambda )$ when $R\rightarrow 0$. Note that (as is well-known [4, 13]) in
the low-velocity limit the SP is proportional to the velocity of particle
(Eqs. (34) and (35)). Thus the vicinage function $\Gamma (\lambda
,R,\vartheta )$ at $\lambda \ll 1$ depends only on interionic distance $R$
and orientation angle $\vartheta $.

\section{NUMERICAL RESULTS AND DISCUSSION}

Using the theoretical results of Secs. 2 and 3, we have made extensive
numerical calculations of stopping power (SP) and related quantities. In
this section we present detailed numerical results for two target materials,
Al and Cu. These two targets have been chosen because of their frequent use
in experiments and also because of their different electron densities. In
our calculations $\chi $ (or $r_s$) is a measure of electron density.

As a simple but generic example of a projectile, we have considered a
diproton cluster for which we present theoretical results for the following
quantities of physical interest: stopping power (SP/2), vicinage function
(VF), angle-averaged stopping power (ASP/2), angle-averaged vicinage
function (AVF) together with the dependence of SP/2 and VF on $R$, the
inter-ionic separation distance within the cluster. The reason why SP has
been divided by a factor of 2 is that the SP results for a diproton cluster
are expected to reduce asymptotically (as $R$ tends to infinity) to those
for two uncorrelated protons, the latter being referred to as ISP. ASP has
been treated in the same way.

In our calculations of these quantities we have employed the linear response
approach which assumes a swift ion-cluster projectile and also that the ion
cluster presents a weak perturbation on the target plasma. The validity of
the linear response approach to study ion-cluster stopping has been
discussed in detail by Zwicknagel and Deutsch [17]. We refer the reader to
their insightful discussion.

We model the Al and Cu targets by a dense (degenerate) electron gas
neutralized by a positive background (the jellium model) with electron
densities appropriate for the respective targets. The linear response of the
target electron gas, which couples the cluster projectile to the target, is
considered at three levels of approximations to the dielectric function $%
\varepsilon (k,\omega )$ as discussed in sections 2 and 3. In the context of
stopping power these approximations are subject to the following general
remarks: The plasmon-pole approximation (PPA) is valid only in the high
velocity regime when the mean velocity $V$ of the cluster is $>v_F$, the
Fermi velocity of the target electrons. For $V<v_F$ and for velocities near
the threshold of collective mode excitations, this approximation is not
adequate. The RPA overcomes this limitation although it cannot account for
short-range correlations in the electron gas. Within PPA itself, PPA-1
(without plasmon dispersion) is more limited than PPA-2 (with plasmon
dispersion). The figures we present serve as a comparative study of how
these levels of approximation affect the various physical quantities related
to stopping power.

Figs. 1-4 show cluster stopping power (CSP and its dependence on various
quantities of experimental interest). Let us first note that these figures
are presented for two specific values, $0$ and $\pi /2$, of the angle $%
\vartheta $. Correlations between the two ions in the dicluster are maximum
and minimum, respectively, for these two values of $\vartheta $. The
objective is then to see how, for these maximum and minimum configurations,
CSP depends on $R$ and $V/v_F$. Fig. 1 shows CSP for Al target with $%
R=10^{-8}{\rm cm}$, as a function of $V/v_F$ for the two above-mentioned
values of $\vartheta $, within PPA. The lines without circles correspond to
PPA-1 and, with circles, to PPA-2. The angular dependence of CSP is
particularly noteworthy. It is seen that in a medium velocity range ($V<2v_F$%
), CSP has a remarkably higher value for the larger value of $\vartheta $.
This is likely due to single-particle excitations in this velocity range. In
the higher velocity range, the dicluster wake-field excitations become
important and we find that the situation is reverse in the higher velocity
range ($V>2v_F$) for which CSP for $\vartheta =0$ is larger than for $%
\vartheta =\pi /2$.

In the low velocity range the difference between PPA-1 and PPA-2 (for both $%
\vartheta =0$ and $\pi /2$) is noticeable while in the high velocity range
this difference becomes negligible. This is again due to single-particle
excitations in the low velocity range. For comparison, we have also
presented the uncorrelated stopping power (ISP).

When we increase the inter-ionic separation distance $R$ from $10^{-8}{\rm cm%
}$ to $5\times 10^{-8}{\rm cm}$, keeping other physical parameters the same,
some interesting changes occur, as can be seen from Fig. 2. A noticeable
change is that now, for $V<2v_F$, CSP for $\vartheta =0$ is higher than that
for $\vartheta =\pi /2$. This sensitivity of CSP to the angle $\vartheta $
as $R$ is varied may be due a combination of factors. The dicluster behaves
like a compact project for small $R$, and like an extended projectile for
large $R$. This has a bearing on $S_{{\rm corr}}$ given in Eqs. (13) and
(22). Correlation effects are expected to be maximum when the two ions are
aligned with each other in the direction of propagation of the dicluster
projectile motion ($\vartheta =0$) while they decay (at least for $V>v_F$)
when $\vartheta $ tends to $\pi /2$, the latter behavior being related to
the wake-field due to the leading ion. The oscillation amplitude in $S_{{\rm %
corr}}$ tends to decrease from $\vartheta =0$ to $\vartheta =\pi /2$ (the \v
{C}herenkov cone). However when $R$ is small each ion is influenced by the
unscreened field of the other ion. For model solid targets, the \v
{C}herenkov cone semivertex is $\vartheta _C={\rm arcsin}(\sqrt{0.6}v_F/V)$
[18]. $\vartheta _C$ approximately equal to 22.8$^0$, 7.4$^0$, and 0.08$^0$
for $V=2v_F$, $V=6v_F$, and $V=10v_F$, respectively. Consequently in the
high velocity range the trailing ion moves inside the \v {C}herenkov cone of
the leading ion only for almost aligned diclusters. The behavior of CSP
shown in Figs. 1 and 2 reflects these features within the linear response
and for PPA. It will be noted that the high values of SP are due to the
PPA-1 approximation. PPA-2 decreases these values to a small extent. Later,
when we use a more realistic, namely RPA, for the linear response function
(Figs. 20 and 21) SP considerably decreases in strength.

Figs. 3 and 4 show SP for Cu, another commonly used metallic target. These
figures show patterns similar to those in Figs. 1 and 2, except that CSP and
ISP have lower values over the entire range of $V/v_F$. This is because Al
has a higher electron density than Cu.

The vicinage function (VF) given by Eq. (8) has been plotted as a function
of the beam velocity for Al target in Figs. 5 and 6. This function shows an
interplay between $\vartheta $ and $R$ more strikingly than CSP. Figs. 7 and
8 display a similar behavior for Cu target.

As stated in Sec. 2, an average stopping power (ASP) is of experimental
interest. Figs. 9 and 10 show ASP for Al and Cu, respectively. ISP is also
shown, for comparison. The role of PPA-1 and PPA-2 is now more clearly seen.

In the same spirit we have plotted AVF for Al and Cu in Figs. 11 and 12. The
role of $R$ is highlighted in these figures. However it will be noticed that
PPA-1 and PPA-2 make practically no distinction for AVF.

We have so far plotted SP or ASP (divided by a factor of 2 in both the
cases) vs the beam velocity $V/v_F$, for some values of the separation
distance $R$. We now look for some complementary information about SP, and
plot SP as a function of $R$ with $V=3v_F$, for Al target. Fig. 13 shows an
oscillatory character of SP with respect to $R$. The oscillations are the
highest for $\vartheta =0$ and lowest for $\vartheta =\pi /2$. The role of
PPA-1 and PPA-2 is clearly seen for $\vartheta =0$. Fig. 14 shows a similar
behavior of SP for Cu although the amplitudes are now weaker.

In the same way the vicinage function (VF) is plotted in Figs. 15 and 16,
for Al and Cu targets, respectively.

Fig. 17 shows ASP vs $R$ for both Al and Cu targets. The difference between
PPA-1 and PPA-2 is negligible and the Cu target has ASP smaller than for Al.
Now, there is something interesting about Fig.18 which shows AVF. The
difference between PPA-1 and PPA-2 is again negligible. But let us note that
data for both Al and Cu lie practically on the same curve! Recalling the
definition of VF, Eq. (8) one can see from Eqs. (12), (16), (21) and (26)
that AVF has a weak dependence on target density. Also, when $\lambda
=V/v_F>2$, $S_{{\rm ind}}$ does not noticeably depend on PPA-1 and PPA-2.
These features combine to lead to the behavior of AVF as seen in Fig. 18.

We have so far presented results for PPA-1 and PPA-2. A more realistic
linear response function, namely the exact random phase approximation (RPA)
will now be used for the metallic target. The theoretical results for SP
etc. have been presented in Sec. 3.3. As part of calculating SP in RPA it is
useful to examine the plasmon dispersion obtained through $\varepsilon
(z_r,u)=0$, where $z$ and $u$ have been defined in Sec. 3.3. Fig. 19
displays $z_r(\chi ,u)$ vs $u$ for three electron density parameter values.

Next we present ISP and CSP in RPA, for Al and Cu in Figs. 20 and 21,
corresponding to $R=10^{-8}{\rm cm}$ and $R=5\times 10^{-8}{\rm cm}$,
respectively. For the sake of a better presentation of the data we have also
separately displayed the Al data in Figs. 20a and 21a. The RPA results show
that SP and ISP decrease in strength with an improved linear response
function. This should be of relevance to experiments.

Next, VF in RPA vs $V/v_F$ is presented in Fig. 22, for Al (lines without
circles) and for Cu (lines with circles), corresponding to $R=10^{-8}{\rm cm}
$ and for $\vartheta =0$ and $\pi /2$. This figure may be compared with
Figs. 5 and 7. For $V/v_F<2$, the curves for VF in RPA tend toward finite
values whereas the VF-curves in PPA do not although the angular trend is
similar. A similar contrast may be noted between Figs. 6 and 8, and Fig. 23,
corresponding to $R=5\times 10^{-8}{\rm cm}$. Again, these findings are of
experimental relevance.

Averaged SP vs. $V/v_F$ in RPA is presented in Fig. 24, for Al (curves
without circles) and for Cu (curves with circles) along with ISP,
corresponding to $R=10^{-8}{\rm cm}$ and $5\times 10^{-8}{\rm cm}$. This
figure may be compared with Figs. 9 and 10. Fig. 24 shows an expected
overall decrease in the strength of ASP in RPA.

There are similarities but also some interesting differences if we compare
Figs. 11 and 12 with Fig. 25. The latter shows AVF in RPA for Al (curves
without circles) and for Cu (curves with circles). The differences are more
noteworthy for $R=10^{-8}{\rm cm}$.

SP vs $R$ in RPA is plotted in Fig. 26, corresponding to $V=3v_F$, for Al
and Cu in the previously stated scheme. When Fig. 26 is compared with Figs.
13 and 14, the differences between PPA and RPA become particularly striking.

A similar contrast is provided by a comparison of Fig. 27 with Figs. 15 and
16, for VF vs. $R$ in RPA and PPA, corresponding to $V=3v_F$ and for $%
\vartheta =0$ and $\pi /2$. For comparison AVF is also plotted in Fig.27.

This completes our extensive presentation of figures exhibiting various
aspects of the stopping power of a diproton cluster in PPA and RPA, for Al
and Cu targets.

\section{SUMMARY}

In this paper we have presented a comprehensive theoretical study of
stopping power (SP) of a dicluster of protons in a metallic target. After a
general introduction to SP of a cluster of two point-like ions, in Sec. 2,
theoretical calculations of SP based on the linear response theory and using
PPA without and with plasmon dispersion and then with RPA are discussed in
Sec. 3. The theoretical expressions for a number of physical quantities
derived in section lead to a detailed presentation, in Sec. 4, of a large
collection of data through figures on correlated stopping power (CSP),
vicinage function (VF), average stopping power (ASP) and average vicinage
function (AVF) of a diproton cluster projectile for two metallic targets, Al
and Cu. Whenever relevant, we have also provided a plot of independent (i.e.
single-ion) stopping power (ISP) for comparison.

With the proviso stated in Sec. 4, SP and related quantities have been
studied within a linear response formalism; some analytical and all
numerical results have been obtained corresponding to three approximations
to the dielectric function of the target electron gas-the plasmon-pole
approximation (PPA) without dispersion (PPA-1) and with dispersion (PPA-2),
and also with the random phase approximation (RPA). To our knowledge this is
the most comprehensive calculation of the SP-related physical quantities
using all the three dominant approximations to the linear response function.
The results we have presented demonstrate that with regard to several
physical quantities of primary interest the difference between PPA and RPA
is substantial while for others, specially for average quantities, this
difference may not be of practical significance.

It will be of interest to go beyond RPA in order to include some short-range
correlations in the electron gas and to study how dicluster SP is affected.
However calculating the linear response function by including electron
energy bands is rather involved and detailed theoretical studies of SP with
band structure effects included have not yet been reported in the
literature. One can include some aspect of band structure in a rather
approximate manner through an effective mass for the electrons.

Another aspect we have not considered in this paper is some effect of
disorder in the target medium. In real metals electrons suffer collisions
with impurities etc. We intend to address this issue in the context of
stopping power in a separate study.

\vspace{2cm}

\begin{center}
{\bf ACKNOWLEDGMENT}
\end{center}

It is a pleasure to thank to Dr. G. Zwicknagel for useful discussions. We
are grateful to V. Nikoghosyan for technical assistance.

\begin{center}
\newpage {\bf Figure Captions}
\end{center}

Fig. 1. SP/2 of a diproton cluster with $R=10^{-8}{\rm cm}$, vs $V/v_F$ for
Al target ($r_s=2.07$). $\vartheta =0$ (dotted line), $\vartheta =\pi /2$
(dashed line); ISP (solid line). The lines with and without circles
correspond to PPA with and without dispersion, respectively.

Fig. 2. SP/2 of a diproton cluster with $R=5\times 10^{-8}{\rm cm}$, vs $%
V/v_F$ for Al target. $\vartheta =0$ (dotted line), $\vartheta =\pi /2$
(dashed line); ISP (solid line). The lines with and without circles
correspond to PPA with and without dispersion, respectively.

Fig. 3. SP/2 of a diproton cluster with $R=10^{-8}{\rm cm}$, vs $V/v_F$ for
Cu target ($r_s=2.68$). $\vartheta =0$ (dotted line), $\vartheta =\pi /2$
(dashed line); ISP (solid line). The lines with and without circles
correspond to PPA with and without dispersion, respectively.

Fig. 4. SP/2 of a diproton cluster with $R=5\times 10^{-8}{\rm cm}$, vs $%
V/v_F$ for Cu target. $\vartheta =0$ (dotted line), $\vartheta =\pi /2$
(dashed line); ISP (solid line). The lines with and without circles
correspond to PPA with and without dispersion, respectively.

Fig. 5. VF of a diproton cluster with $R=10^{-8}{\rm cm}$, vs $V/v_F$ for Al
target. $\vartheta =0$ (solid line), $\vartheta =\pi /2$ (dotted line). The
lines with and without circles correspond to PPA with and without
dispersion, respectively.

Fig. 6. VF of a diproton cluster with $R=5\times 10^{-8}{\rm cm}$, vs $V/v_F$
for Al target. $\vartheta =0$ (solid line), $\vartheta =\pi /2$ (dotted
line). The lines with and without circles correspond to PPA with and without
dispersion, respectively.

Fig. 7. VF of a diproton cluster with $R=10^{-8}{\rm cm}$, vs $V/v_F$ for Cu
target. $\vartheta =0$ (solid line), $\vartheta =\pi /2$ (dotted line). The
lines with and without circles correspond to PPA with and without
dispersion, respectively.

Fig. 8. VF of a diproton cluster with $R=5\times 10^{-8}{\rm cm}$, vs $V/v_F$
for Cu target. $\vartheta =0$ (solid line), $\vartheta =\pi /2$ (dotted
line). The lines with and without circles correspond to PPA with and without
dispersion, respectively.

Fig. 9. ASP/2 of a diproton cluster with $R=10^{-8}{\rm cm}$ (dotted line)
and $R=5\times 10^{-8}{\rm cm}$ (dashed line) vs $V/v_F$ for Al target.
Solid line, ISP. The lines with and without circles correspond to PPA with
and without dispersion, respectively.

Fig. 10. ASP/2 of a diproton cluster with $R=10^{-8}{\rm cm}$ (dotted line)
and $R=5\times 10^{-8}{\rm cm}$ (dashed line) vs $V/v_F$ for Cu target.
Solid line, ISP. The lines with and without circles correspond to PPA with
and without dispersion, respectively.

Fig. 11. AVF of a diproton cluster with $R=10^{-8}{\rm cm}$ (solid line) and 
$R=5\times 10^{-8}{\rm cm}$ (dotted line) vs $V/v_F$ for Al target. The
lines with and without circles correspond to PPA with and without
dispersion, respectively.

Fig. 12. AVF of a diproton cluster with $R=10^{-8}{\rm cm}$ (solid line) and 
$R=5\times 10^{-8}{\rm cm}$ (dotted line) vs $V/v_F$ for Cu target. The
lines with and without circles correspond to PPA with and without
dispersion, respectively.

Fig. 13. SP/2 of a diproton cluster with $V=3v_F$ vs $R$ for Al target. $%
\vartheta =0$ (solid line), $\vartheta =\pi /2$ (dotted line). The lines
with and without circles correspond to PPA with and without dispersion,
respectively.

Fig. 14. SP/2 of a diproton cluster with $V=3v_F$ vs $R$ for Cu target. $%
\vartheta =0$ (solid line), $\vartheta =\pi /2$ (dotted line). The lines
with and without circles correspond to PPA with and without dispersion,
respectively.

Fig. 15. VF of a diproton cluster with $V=3v_F$ vs $R$ for Al target. $%
\vartheta =0$ (solid line), $\vartheta =\pi /2$ (dotted line). The lines
with and without circles correspond to PPA with and without dispersion,
respectively.

Fig. 16. VF of a diproton cluster with $V=3v_F$ vs $R$ for Cu target. $%
\vartheta =0$ (solid line), $\vartheta =\pi /2$ (dotted line). The lines
with and without circles correspond to PPA with and without dispersion,
respectively.

Fig. 17. ASP/2 of a diproton cluster with $V=3v_F$ vs $R$ for Al (the lines
with square symbols) and Cu (the lines with circles) targets. Dotted and
solid lines, PPA with and without dispersion, respectively.

Fig. 18. AVF of a diproton cluster with $V=3v_F$ vs $R$ for Al (the lines
with square symbols) and Cu (the lines with circles) targets. Dotted and
solid lines, PPA with and without dispersion, respectively.

Fig. 19. Relation between $z_r(\chi ,u)$ and $u$, as obtained from the
dispersion equation $\varepsilon \left( z_r,u\right) =0$. Solid line: $\chi
=0.5$, dashed line: $\chi =0.15$, dotted line: $\chi =0.05$.

Fig. 20. SP/2 of a diproton cluster with $R=10^{-8}{\rm cm}$ in a RPA vs $%
V/v_F$ for Al (the lines without symbols) and Cu (the lines with circles)
targets. $\vartheta =0$ (dotted line), $\vartheta =\pi /2$ (dashed line);
ISP (solid line).

Fig. 20a. SP/2 of a diproton cluster with $R=10^{-8}{\rm cm}$ in a RPA vs $%
V/v_F$ for Al target. $\vartheta =0$ (dotted line), $\vartheta =\pi /2$
(dashed line); ISP (solid line).

Fig. 21. SP/2 of a diproton cluster with $R=5\times 10^{-8}{\rm cm}$ in a
RPA vs $V/v_F$ for Al (the lines without symbols) and Cu (the lines with
circles) targets. $\vartheta =0$ (dotted line), $\vartheta =\pi /2$ (dashed
line); ISP (solid line).

Fig. 21a. SP/2 of a diproton cluster with $R=5\times 10^{-8}{\rm cm}$ in a
RPA vs $V/v_F$ for Al target. $\vartheta =0$ (dotted line), $\vartheta =\pi
/2$ (dashed line); ISP (solid line).

Fig. 22. VF of a diproton cluster with $R=10^{-8}{\rm cm}$ in a RPA vs $%
V/v_F $ for Al (the lines without symbols) and Cu (the lines with circles)
targets. $\vartheta =0$ (solid line), $\vartheta =\pi /2$ (dotted line).

Fig. 23. VF of a diproton cluster with $R=5\times 10^{-8}{\rm cm}$ in a RPA
vs $V/v_F$ for Al (the lines without symbols) and Cu (the lines with
circles) targets. $\vartheta =0$ (solid line), $\vartheta =\pi /2$ (dotted
line).

Fig. 24. ASP/2 of a diproton cluster with $R=10^{-8}{\rm cm}$ (dotted line)
and $R=5\times 10^{-8}{\rm cm}$ (dashed line) in a RPA vs $V/v_F$ for Al
(the lines without symbols) and Cu (the lines with circles) targets; ISP
(solid line).

Fig. 25. AVF of a diproton cluster with $R=10^{-8}{\rm cm}$ (solid line) and 
$R=5\times 10^{-8}{\rm cm}$ (dotted line) in a RPA vs $V/v_F$ for Al (the
lines without symbols) and Cu (the lines with circles) targets.

Fig. 26. SP/2 of a diproton cluster with $V=3v_F$ in a RPA vs $R$ for Al
(the lines without symbols) and Cu (the lines with circles) targets; ASP
(solid line), $\vartheta =0$ (dotted line), $\vartheta =\pi /2$ (dashed
line).

Fig. 27. VF of a diproton cluster with $V=3v_F$ in a RPA vs $R$ for Al (the
lines without symbols) and Cu (the lines with circles) targets; AVF (solid
line), $\vartheta =0$ (dotted line), $\vartheta =\pi /2$ (dashed line).

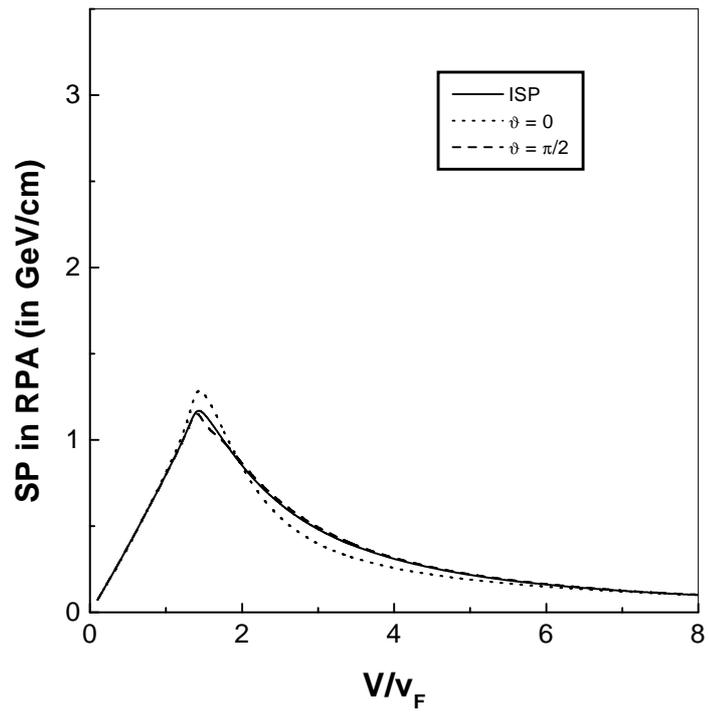
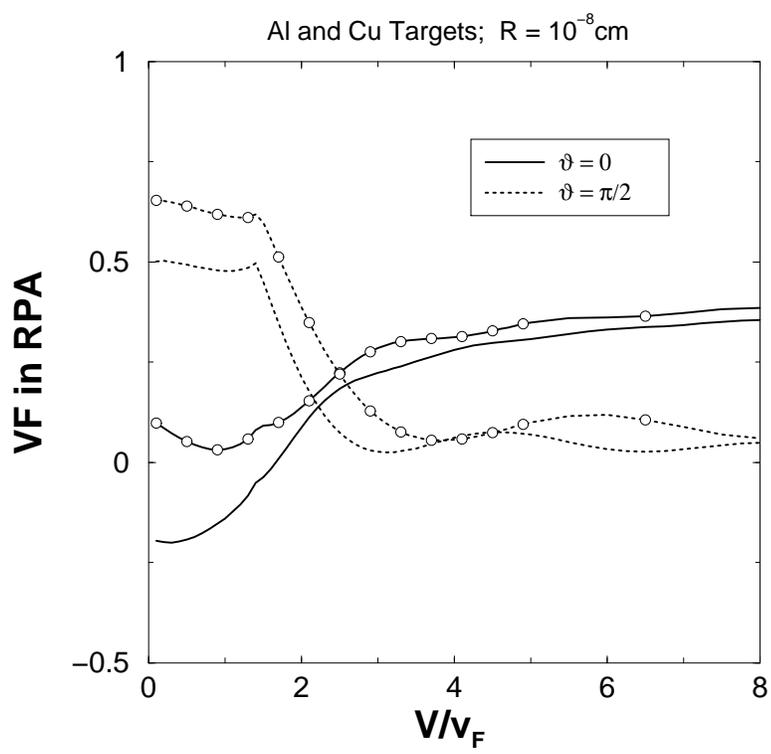
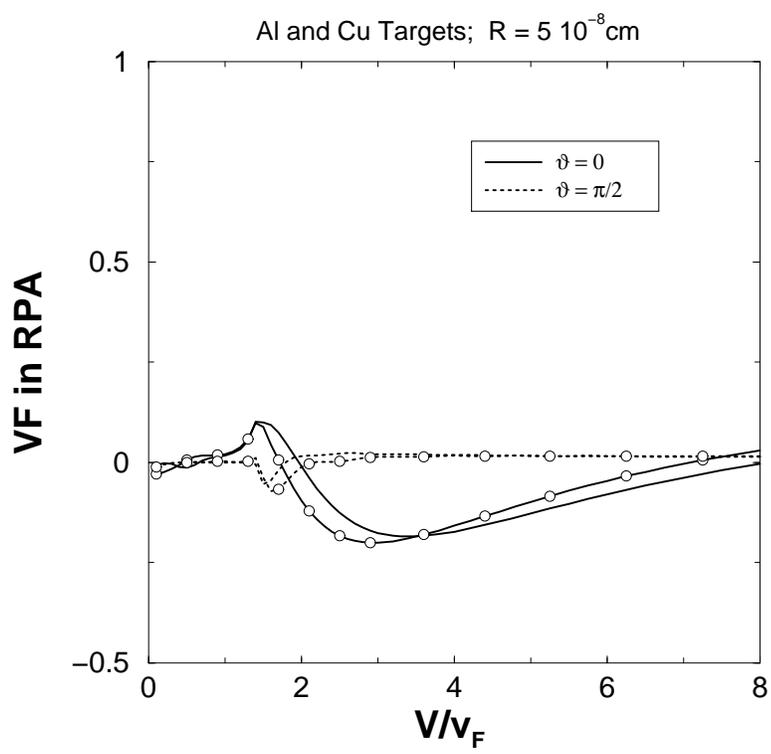
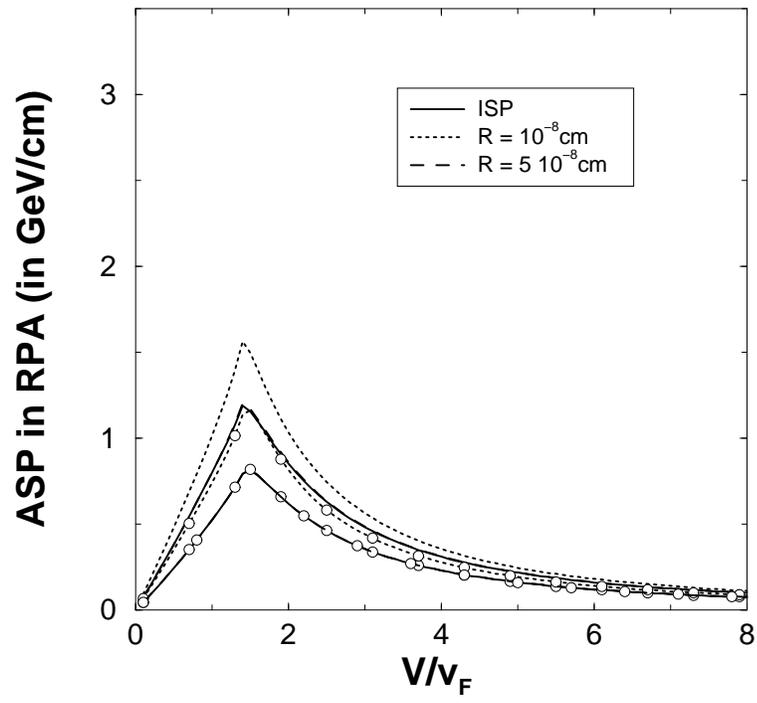
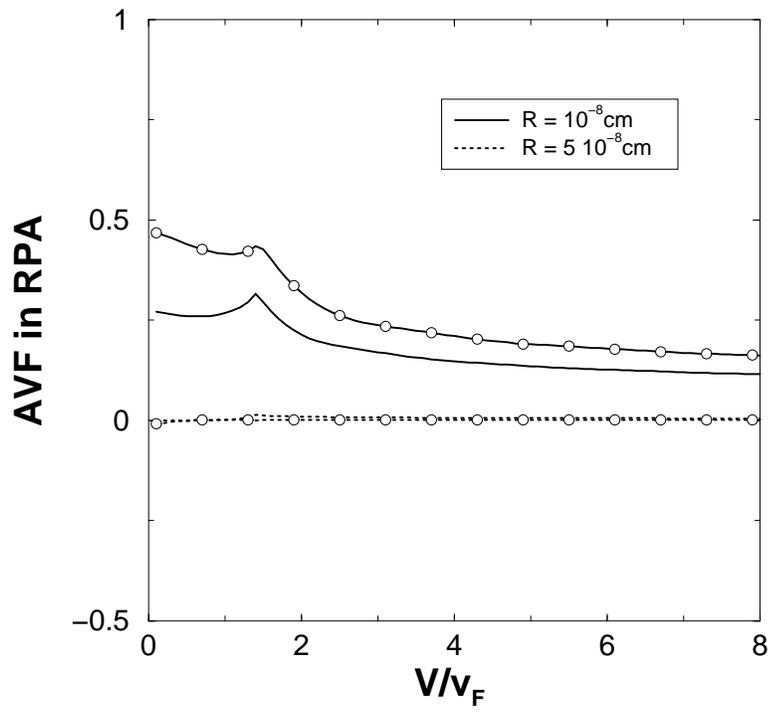
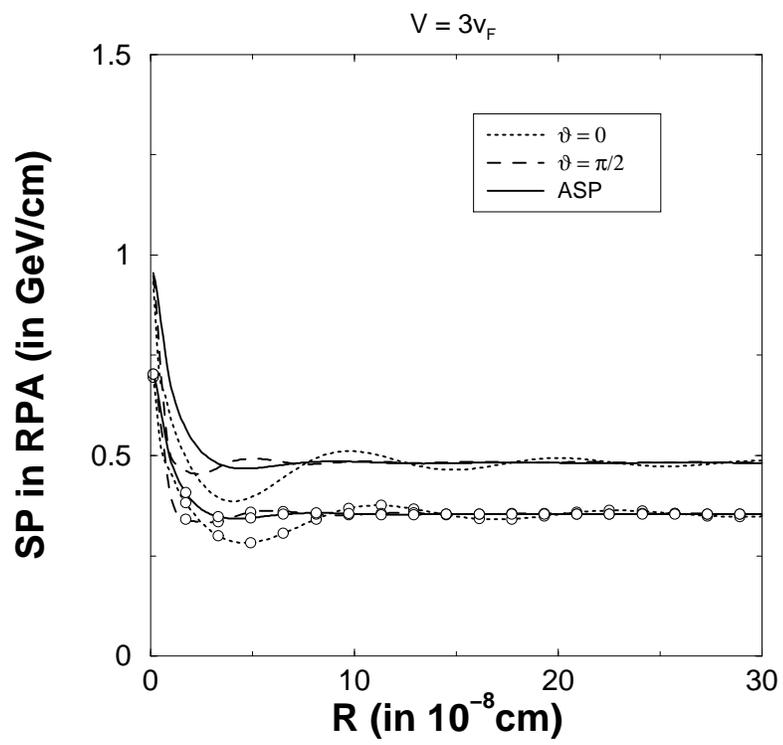
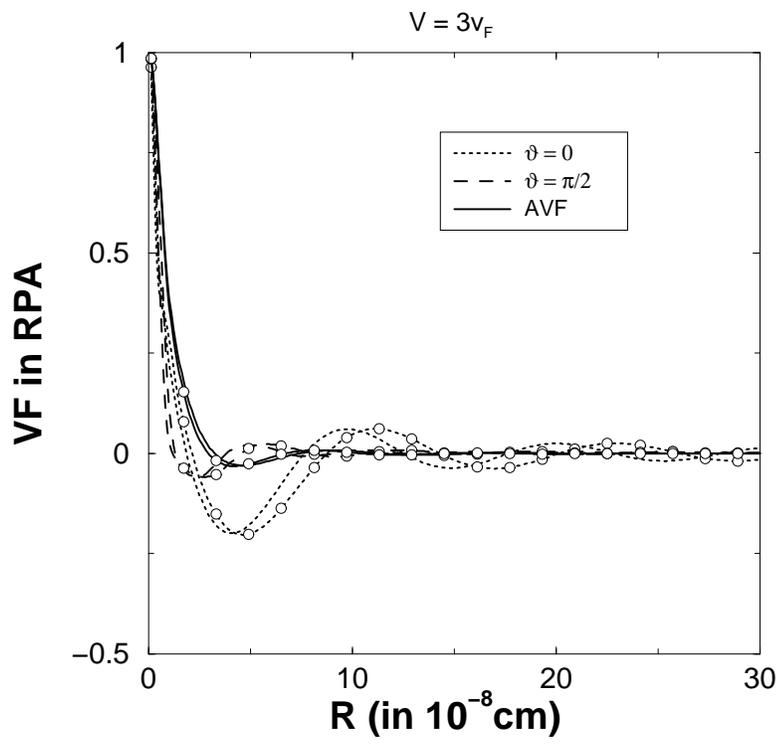
\end{document}